\documentclass[12pt,onecolumn]{revtex4-1}
\usepackage{float}
\usepackage{xcolor}
\usepackage{graphicx}% Include figure files
\usepackage{dcolumn}% Align table columns on decimal point
\usepackage{bm}% bold math
\usepackage{hyperref}% add hypertext capabilities
\usepackage{color}
\usepackage{amsmath}
\usepackage{amsfonts}
\usepackage{amssymb}

\usepackage{tipa}
\usepackage{chemarrow}
\usepackage{braket}

\makeatletter
\renewcommand{\section}{\@startsection{section}{1}{0mm}
  {-\baselineskip}{0.5\baselineskip}{\bf\leftline}}
\makeatother

\begin{document}

%\preprint{APS/123-QED}

\title{Dirac Fermions in Antiferromagnetic Semimetal} %Title of paper

% Force line breaks with \\
%\thanks{A footnote to the article title}%

%\author{xxx}
%\affiliation{yyy}

\author{Peizhe Tang}
\altaffiliation{These authors contribute equally to this work.}
\affiliation{Department of Physics, McCullough Building, Stanford University, Stanford, California 94305-4045, USA}

\author{Quan Zhou}
\altaffiliation{These authors contribute equally to this work.}
\affiliation{Department of Physics, McCullough Building, Stanford University, Stanford, California 94305-4045, USA}

\author{Gang Xu}
\affiliation{Department of Physics, McCullough Building, Stanford University, Stanford, California 94305-4045, USA}

\author{Shou-Cheng Zhang}
\email{Corresponding author: sczhang@stanford.edu}
\affiliation{Department of Physics, McCullough Building, Stanford University, Stanford, California 94305-4045, USA}

\date{\today}% It is always \today, today,
             %  but any date may be explicitly specified

\begin{abstract}
The analogues of elementary particles have been extensively searched for in condensed matter systems because of both scientific interests and technological applications. Recently massless Dirac fermions were found to emerge as low energy excitations in the materials named Dirac semimetals. All the currently known Dirac semimetals are nonmagnetic with both time-reversal symmetry $\mathcal{T}$ and inversion symmetry $\mathcal{P}$. Here we show that Dirac fermions can exist in one type of antiferromagnetic systems, where $\mathcal{T}$ and $\mathcal{P}$ are broken but their combination $\mathcal{PT}$ is respected. We propose orthorhombic antiferromagnet CuMnAs as a candidate, analyze the robustness of the Dirac points with symmetry protections, and demonstrate its distinctive bulk dispersions as well as the corresponding surface states by \emph{ab initio} calculations. Our results give a new route towards the realization of Dirac materials, and provide a possible platform to study the interplay of Dirac fermion physics and magnetism.
\end{abstract}

\maketitle

The great success in the field of topological insulators\cite{Qi-2011,Hasan2010RMP} since last decade inspired the study of topological features of metals. Topological metals have nontrivial surface states and their bulk Fermi surfaces can be topologically characterized\cite{volovik2009universe}. Among them, Dirac semimetals\cite{Na3Bi-Wang-2012,Cd3As2-Wang-2013,BiO2-Young-2012} have received special attention because they host massless Dirac fermions, which are building blocks of the Standard Model. In such Dirac materials, two doubly degenerate bands contact at discrete momentum points called Dirac points (DPs), and disperse linearly along all directions around these points. The fourfold degenerate DPs are unstable by themselves, hence symmetry protection is necessary\cite{yang2014}. Following this guideline, several three-dimensional (3D) Dirac semimetals have been theoretically proposed, and some of them were experimentally verified recently\cite{Na3Bi-liu-2014,Cd3As2-liu2014stable}. All these materials have time-reversal symmetry $\mathcal{T}$, inversion symmetry $\mathcal{P}$, and certain crystalline rotation symmetry.

If some of the symmetries are broken, Dirac fermions can in general be destroyed. For instance, when either $\mathcal{T}$ or $\mathcal{P}$ is broken, each doubly degenerate band is split off, so that the Dirac cones generally split into multiple Weyl cones\cite{murakami2007phase}. This gives birth to Weyl semimetals\cite{Wan-2011,Gangxu2011,Wen-TaAs-2015,huang2015weyl,soluyanov2015type,SunY2015,ruan2015symmetry}, and the chiral-anomaly related phenomena can be observed as a signature\cite{nielsen1983anomaly,xiong2015evidence}. But the result of both $\mathcal{T}$ and $\mathcal{P}$ breaking remains obscure until now. In other words, it is natural to ask whether Dirac fermions can still exist in the absence of both $\mathcal{T}$ and $\mathcal{P}$.

In this letter, we answer the question in the affirmative, and provide a concrete example of such Dirac semimetallic phase. We consider 3D systems with the anti-ferromagnetic (AFM) order that breaks both $\mathcal{T}$ and $\mathcal{P}$ but respects their combination $\mathcal{PT}$. The low energy physics can be explicitly captured by the following four-band effective model
\begin{equation}
H(\bm{k}) = d_0(\bm{k}) \mathbb{I}_{4\times 4} + d_1(\bm{k})\tau_x + d_2(\bm{k})\tau_z + d_3(\bm{k}) \tau_y\sigma_x + d_4(\bm{k})\tau_y\sigma_y + d_5(\bm{k})\tau_y\sigma_z,
\label{equ:1}
\end{equation}
where $d_i(\bm{k}), i = 0, 1, \cdots, 5$ are real functions of momentum $\bm{k}$, and $\tau_{x,y,z}$ ($\sigma_{x,y,z}$) are Pauli matrices for orbital (spin-related AFM) basis. The anti-unitary $\mathcal{PT}$ symmetry satisfying $(\mathcal{PT})^2 = -1$ is given as $\mathcal{PT} = i\sigma_y K$ where $K$ is complex conjugation. Due to this symmetry, the last five terms in $H(\bm{k})$ anti-commute with one another, therefore every band must be doubly degenerate (this degeneracy holds for all $\mathcal{PT}$ invariant systems, see Supplementary Section 1) with energy spectrum
\begin{equation}
\epsilon_{\pm} = d_0(\bm{k}) \pm \sqrt{\sum_{i = 1, \cdots, 5} d^2_i(\bm{k})}.
\end{equation}
If the two doubly degenerate bands cross each other at isolated momentum points, e.g. at $\bm{k} = \bm{q}$ with $d_i(\bm{q}) = 0$ for $i = 1, \cdots, 5$, $\bm{k}=\bm{q}$ must be a fourfold degenerate point, which can be Dirac-like when additional constraints are enforced by crystalline symmetries.

To realize stable fourfold degenerate crossing points, the generic way is to let the bands carry different representations of certain symmetries in the system\cite{DNL-FuLiang-2015,Young-2015}. For our AFM model, however, there is a simpler starting point for the investigation. Suppose the local magnetic moments are along $z$ direction. When spin-orbit coupling (SOC) effect is ignored, $d_3(\bm{k})$ and $d_4(\bm{k})$ vanish because they correspond to spin-flip processes. Thus the crossing points exist in general, because with three momentum components one can tune $d_1(\bm{k}), d_2(\bm{k})$ and $d_5(\bm{k})$ to zero simultaneously. The crossing points can be DPs in the presence of crystalline symmetry, and they might be protected by the symmetry even if SOC is included.

In accordance with our analysis, we discover that the AFM semimetals, orthorhombic CuMnAs and CuMnP (see Supplementary Section 5), can host the Dirac fermions around the Fermi level. As room temperature anti-ferromagnets, these materials have been synthesized and studied previously for the potential applications in spintronics\cite{maca2012room,wadley2015electrical}. The crystal structure has the nonsymmorphic space group D$_{2h}$ (Pnma) with four formula units in the primitive unit cell, (see Fig. 1(a), (b) for the structure and the first Brillouin zone (BZ)). The space group consists of eight symmetry operations that can be generated by three of them: the inversion $\mathcal{P}$, the gliding mirror reflection of $y$ plane $R_y = \{m_y | (0, \frac{1}{2}, 0)\}$, and the twofold screw rotation along $z$ axis $S_{2z} = \{C_{2z} | (\frac{1}{2}, 0, \frac{1}{2})\}$ (see Fig. 1(c)). The two nonsymmorphic symmetries $R_y$ and $S_{2z}$ are important in our symmetry arguments (see Supplementary Section 2).

Nonzero magnetic moments on Mn atoms with 3$d$ electrons exist in CuMnAs and CuMnP, and they tend to order anti-ferromagnetically (see Supplementary Section 4). The magnetic structure typically breaks some symmetries from the original space group. For the most energy-favored AFM configuration shown in Fig. 1(a), the magnetic moments on the inversion-related Mn atoms are aligned along opposite directions, therefore both $\mathcal{T}$ and $\mathcal{P}$ are broken whereas $\mathcal{PT}$ still holds. If SOC is absent, spin internal space is decoupled from real space, so the spatial symmetries $R_y$ and $S_{2z}$ are kept. When SOC is included, however, residual symmetries depend on the orientation of magnetic moments, for example, only $S_{2z}$ will survive if magnetic moments are along $z$ axis, as shown in Fig. 1(d).

With the crystal structure and symmetry operations in mind, we begin to present our results of band structure calculations as well as effective model analysis (see Supplementary Section 9 for details of parameter choices). When SOC is turned off in the AFM system, our first-principle calculations show an elliptical Dirac nodal line (DNL) on the $k_y=0$ plane around the Fermi level, with its center at X point (see Fig. 2(a)). We calculated the band dispersions for various situations (see Supplementary Section 7), and found no gap opening along the nodal structure as long as $R_y$ is present. Nevertheless, because $R_y$ and $\mathcal{PT}$ commute on the $k_y=0$ plane, no rigorous symmetry protection exists for the band crossing here in general sense (see Supplementary Section 2). By checking the orbital composition of the bands, we confirmed that the existence of such DNL in the absence of SOC is associated with the $R_y$ symmetry properties of the underlying atomic orbitals (see Supplementary Section 7). Corresponding to the DNL in the bulk, a nontrivial surface state appears inside the projection of the DNL on the (010) surface (see Fig. 2 (f), (g) and (h)). This dispersive drumhead-like surface state can be detected as a clear signature of the DNL semimetal\cite{DNL-Kane-2015,DNL-HuXiao-2015}.

When we still exclude SOC but break $R_y$ (see Supplementary Section 8), band gap opens along the entire DNL except at four discrete points. One pair of the fourfold degenerate points is located along the high-symmetric X-U direction, and another pair is in the interior of the BZ. We verified the first pair as DPs with linear dispersions shown in Fig. 2(b). The DPs are guaranteed to exist by the screw rotation symmetry $S_{2z}$. Unlike $R_y$, $S_{2z}$ anti-commutes with $\mathcal{PT}$ along the X-U line, so the doubly degenerate states at each $\bm{k}$ point on this line have the same $S_{2z}$ eigenvalues. When two bands with opposite $S_{2z}$ eigenvalues cross each other, the crossing point must be stable. Based on \emph{ab initio} results, we calculated the eigenvalues of $S_{2z}$ of the bands near the Fermi level, and the results match the symmetry argument exactly (see Supplementary Sections 2 and 6).

To check the nature of the DPs, we derive the low energy effective model (see Supplementary Section 3). As we mentioned above, our AFM system without SOC is described by (we ignore the overall shift $\mathbb{I}_{4\times4}$ term in the following)
\begin{equation}
H(\bm{k}) = d_1(\bm{k})\tau_x + d_2(\bm{k})\tau_z + d_5(\bm{k})\tau_y\sigma_z.
\end{equation}
On the high symmetry line X-U, the screw rotation symmetry $S_{2z}$ is represented by $S_{2z} = ie^{-i\frac{k_z}{2}}\tau_z$. Expanding the Hamiltonian around one DP and forcing the symmetry constraints, we can obtain the exact Dirac-type Hamiltonian
\begin{equation}
\mathcal{H}_{Dirac} = (v_{11} k_x + v_{12} k_y)\tau_x + v_{33} k_z \tau_z + (v_{21} k_x + v_{22} k_y) \tau_y\sigma_z,
\end{equation}
where $v_{ij} (i, j = 1, 2, 3)$ are velocity coefficients for different directions. These parameters are obtained from our calculations, and the resulted Dirac cones are anistropic (see Fig. 2(c), (d)). Splitting $\mathcal{H}_{Dirac}$ in two blocks that correspond to $\sigma_z = \pm 1$, we can decouple each Dirac cone into two Weyl cones with opposite chiralities (see Fig. 2 (e))
\begin{equation}
\mathcal{H}^+_{Weyl} = (v_{11} k_x + v_{12} k_y)\tau_x + v_{33} k_z \tau_z + (v_{21} k_x + v_{22} k_y) \tau_y,
\end{equation}
\begin{equation}
\mathcal{H}^-_{Weyl} = (v_{11} k_x + v_{12} k_y)\tau_x + v_{33} k_z \tau_z - (v_{21} k_x + v_{22} k_y) \tau_y.
\end{equation}
Since SOC is absent, the AFM basis $\sigma_z = \pm 1$ is almost equivalent to physical spin basis (see Supplementary Section 1). We thus calculated the surface states on (010) surface for each spin component, as shown in Fig. 2(i-k). It is clear that Fermi arcs exist on the surface, and they connect pairs of Weyl points with opposite chiralities. For either spin component, the chiralities of the Weyl points on the X-U line are found to be the same, therefore it is reasonable that the other two Weyl points that carry opposite chiralities appear in the BZ such that the total chirality vanishes\cite{nielsen1981absence}.

When SOC is turned on, the band structures sensitively depend on the orientation of the local magnetic moments on Mn atoms, as the remaining symmetries are different. If the moments are aligned along $z$ direction (see Fig. 1(a)), only $S_{2z}$ symmetry from the space group survives. In this case, the symmetry argument for the robust crossing points along the X-U line still holds, thus stable fourfold degenerate points protected by $S_{2z}$ symmetry exist along the rotation axis (see Fig. 3(a)). The effective model near each degenerate point is derived in the same way,
\begin{equation}
\mathcal{H} = \mathcal{H}_{Dirac} + (\delta_1 k_x + \delta_2 k_y)\tau_y\sigma_x + (\delta_3 k_x + \delta_4 k_y)\tau_y\sigma_y,
\end{equation}
where the small perturbation terms $\delta_i, (i= 1, 2, 3, 4)$ are purely produced by SOC which can be treated as weak coupling between the two Weyl fermions at each Dirac cone. The calculated electronic structures of AFM CuMnAs are shown in Fig. 3(b-d). It is clear that no gap opens at the crossing point along the X-U line, and that nontrivial surface states appear on the (010) surface which connect two gapless points. However, if the orientation of magnetic moments is along arbitrary directions, $S_{2z}$ is broken and no stable degenerate points are found (see Supplementary Section 8).

Finally we discuss the experimental detection and some new physics of the Dirac fermions in AFM systems. Similar to normal Dirac and Weyl semimetals, the nontrivial surface state and the orbital texture of Dirac cones could be measured by angle-resolved photoemission spectroscopy and would be direct evidence for the Dirac fermions\cite{Na3Bi-xu-2015,xu2015discovery}. Furthermore, magneto-transport experiments can be taken to identify chiral anomaly in the Dirac systems\cite{xiong2015evidence}, though it might be more complicated here because two Weyl cones at the same DPs are intrinsically coupled. In addition, AFM fluctuations exist generically in CuMnAs and CuMnP. They can be regarded as dynamical axion field\cite{li2010dynamical} when the anti-ferromagnets are made fully insulating under certain conditions. In our case, the fluctuations not only couple to the Dirac fermions, but also affect crystalline symmetries. The exact description of interplay between Dirac fermions, the AFM fluctuations, and the symmetry breaking at the moment remains an open question.

The first principle calculations were carried out by density functional theory method with the projector augmented wave method\cite{Bloch1994}, as implemented in the Vienna \textit{ab initio} simulation package\cite{Kresse1996}. The Perdew-Burke-Ernzerhof exchange-correlation functional and the plane wave basis with energy cutoff of $\mathrm{300~eV}$ were employed. The structure was allowed to be fully relaxed until the residual forces are less than $1\times 10^{-3}~\mathrm{eV/\AA}$ . And the relaxed lattice parameters (see Fig. 1(a)) were $a=6.577~\mathrm{\AA}$, $b=3.854~\mathrm{\AA}$, and $c=7.310~\mathrm{\AA}$ for orthorhombic CuMnAs, and $a=6.318~\mathrm{\AA}$, $b=3.723~\mathrm{\AA}$, and $c=7.088~\mathrm{\AA}$ for orthorhombic CuMnP respectively. The Monkhorst-Pack $k$ points were 9$\times$15$\times$9, and SOC was included in self-consistent electronic structure calculations. The maximally localized Wannier functions\cite{mostofi2008wannier90} were constructed to obtain the tight-binding Hamiltonian, which is used to calculate the bulk Fermi surface, surface electronic spectrum and surface states.

\bibliography{references}

\begin{itemize}
\item We acknowledge the Department of Energy, Office of Basic Energy Sciences, Division of Materials Sciences and Engineering, under contract DE-AC02-76SF00515, and FAME, one of six centers of STARnet, a Semiconductor Research Corporation program sponsored by MARCO and DARPA.
 \item P.T., Q.Z., G.X., and S.-C.Z. conceived and designed the project. P.T. performed the first principles calculations, Q.Z. performed theoretical analysis, P.T. and Q.Z. analyzed the data and wrote the manuscript. All authors commented on the manuscript. P.T. and Q.Z. contributed equally to this work.
 \item The authors declare that they have no competing financial interests.
 \item Correspondence should be addressed to S.-C. Z.
\end{itemize}

\begin{figure}
\centering
\begin{minipage}[t]{1.0\textwidth}
\includegraphics[width=\textwidth]{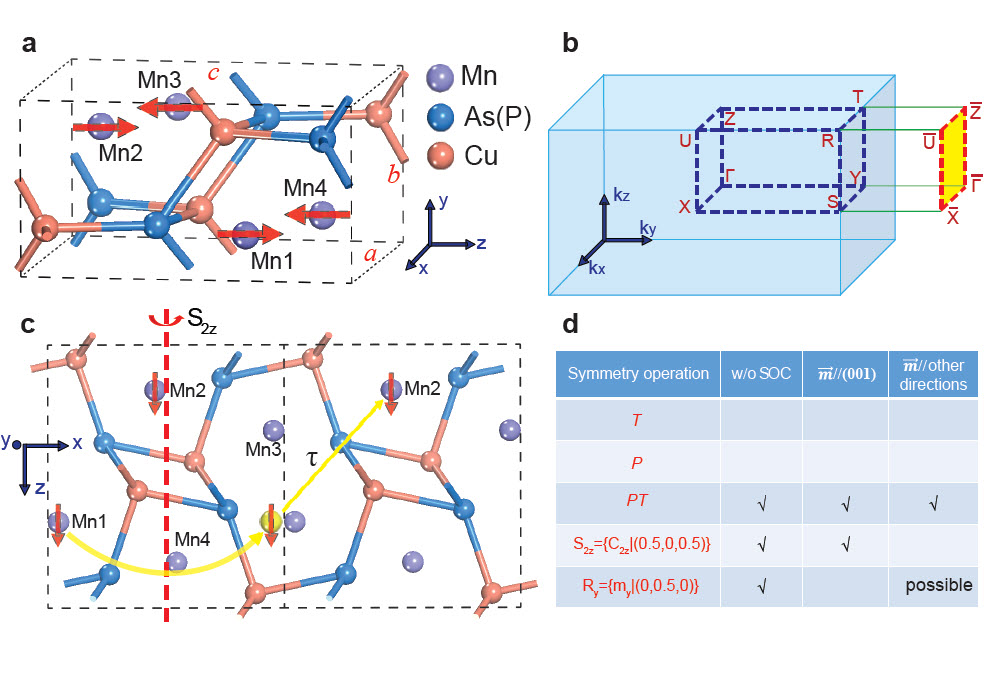}
\end{minipage}
\caption{(Color online) \textbf{The crystal structure, Brillouin zone and crystal symmetry of orthorhombic CuMnAs.} (a) The crystal structure of the orthorhombic CuMnAs(P). The red arrows stand for the orientations of the magnetic moments on Mn atoms. \emph{a}, \emph{b} and \emph{c} denote three primitive lattice vectors. (b) The Brillouin zone and its projection to the (010) surface. (c) Illustration of the screw rotation symmetry $S_{2z} = \{C_{2z} | (\frac{1}{2}, 0, \frac{1}{2})\}$. The red dashed line stands for the rotation axis. The yellow ball represents the position of the Mn atom after $C_{2z}$ rotation, and $\vec{\tau}=(\frac{1}{2}, 0, \frac{1}{2})$ is the half translation along (101) direction. (d) Symmetries for different magnetic configurations with and without SOC. $\vec{m}$ stands for the orientation of the magnetic moments on Mn atoms when SOC is considered.}
\label{fig:cry}
\end{figure}

\begin{figure}
\centering
\begin{minipage}[t]{0.85\textwidth}
\includegraphics[width=\textwidth]{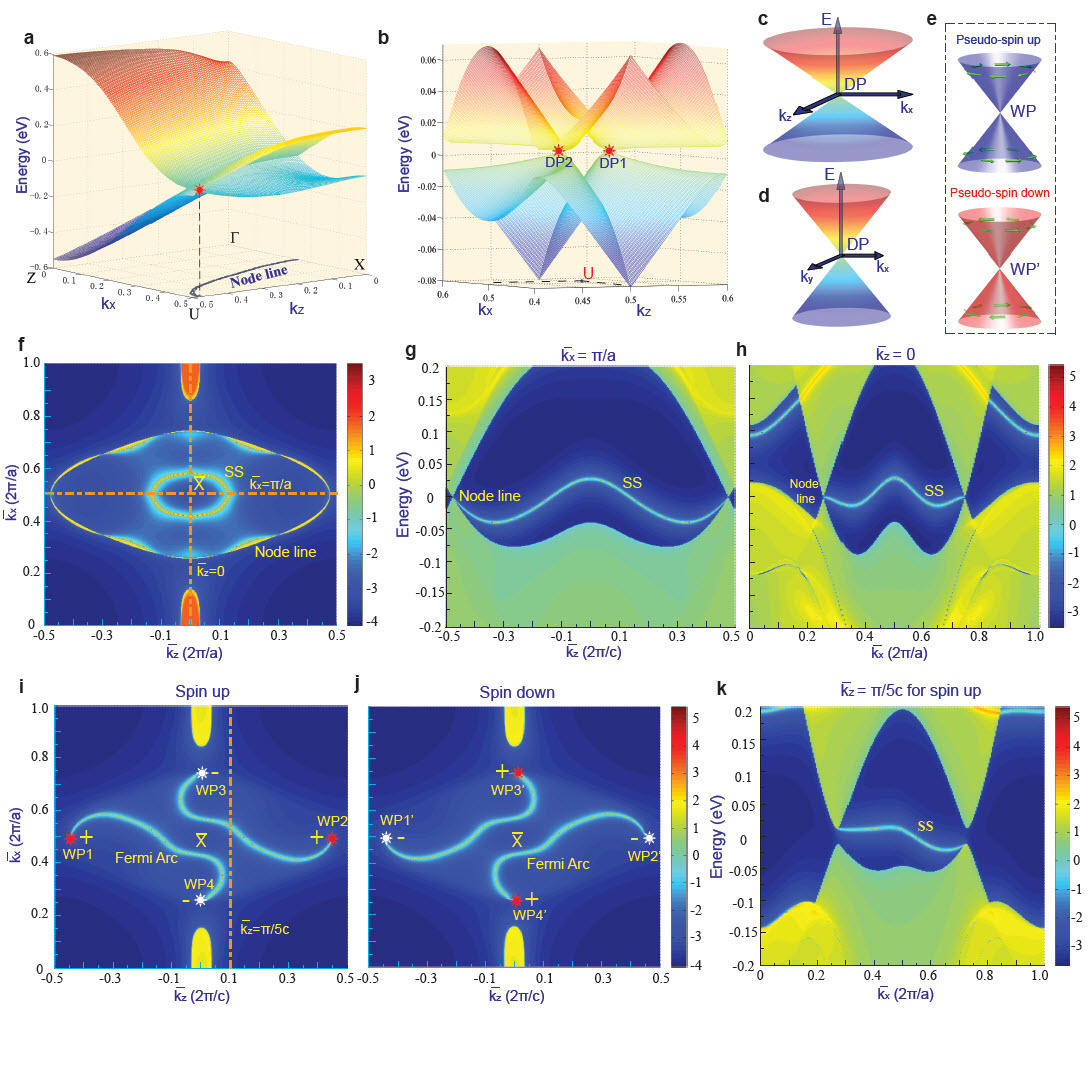}
\end{minipage}
\caption{(Color online) \textbf{Electronic structures of CuMnAs without SOC} (a) The electronic structure of orthorhombic CuMnAs around the Fermi level when $R_y$ is present. The black line represents the Dirac node line. (b) The electronic structure of orthorhombic CuMnAs when $R_y$ is broken by shear strain. The red stars stand for the Dirac points (DPs) around the Fermi level. (c-d) Schematics of the projected anisotropic Dirac cone into the (\emph{k$_{x}$},\emph{k$_{z}$},\emph{E}) and (\emph{k$_{x}$},\emph{k$_{y}$},\emph{E}) space reconstructed from the fitting parameters. (e) Schematics of the degenerate Weyl points (WPs) with different pseudo-spin $\sigma_z = \pm 1$. The green arrows represent the orbital texture. (f) The Fermi surface contour on the (010) surface when $R_y$ is present, and (g-h) the corresponding electronic spectra along $\bar{k_{x}}$=$\pi/a$ and $\bar{k_{z}}$=0 when $R_y$ is present. (i-j) The Fermi surface contours on the (010) surface for spin up and down states respectively when $R_{y}$ is broken. The red and white stars represent the WPs with different topological charges. (k) The corresponding electronic spectrum for spin up states along $\bar{k_{z}}$=$\pi/5c$. The Fermi level is set to zero.}
\label{fig:2}
\end{figure}

\begin{figure}
\centering
\begin{minipage}[t]{0.8\textwidth}
\includegraphics[width=\textwidth]{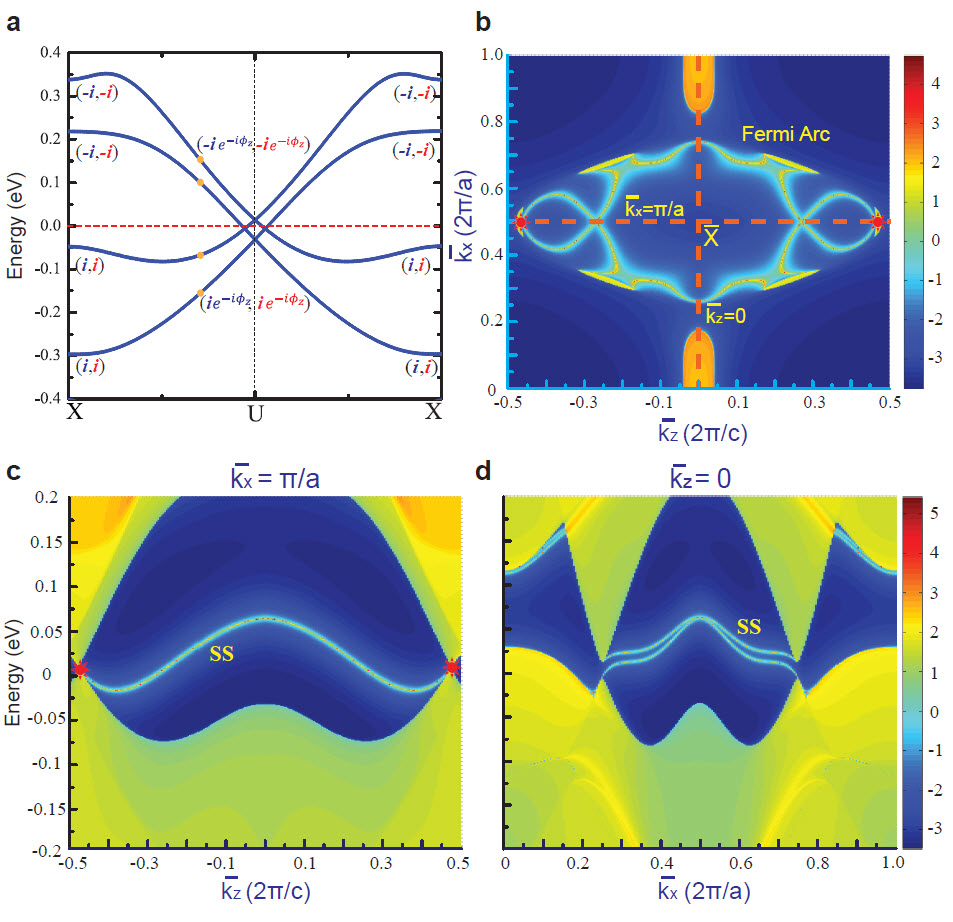}
\end{minipage}
\caption{(Color online) \textbf{Electronic structures of CuMnAs with SOC.} (a) The band structure of orthorhombic CuMnAs with SOC along X-U-X line when the orientation of magnetic moments is along \emph{z}-direction. $\pm i$ represent the eigenvalues of screw rotation symmetry $S_{2z}$ at the X point. Along the line, the eigenvalue of $S_{2z}$ is $\pm ie^{-i\phi_{z}}$. The blue and red colors stand for different spin states. (b) The Fermi surface contour on the (010) surface. (c-d) The corresponding electronic spectra along $\bar{k_{x}}$=$\pi/a$ and $\bar{k_{z}}$=0. The red stars stand for the gapless point protected by the screw symmetry $S_{2z}$. The Fermi level is set to zero.}
\label{fig:3}
\end{figure}

\end{document}